\documentclass[10pt]{iopart}
\usepackage{todonotes}
\usepackage{subfig}
\usepackage{hyperref}
\usepackage{cite}
\usepackage{amsmath,amssymb,amsfonts}
\usepackage{graphicx}
\usepackage{textcomp}
\usepackage{booktabs}
\usepackage{multirow}
\usepackage{adjustbox}

\def\etal{\emph{et al.}}

\usepackage{orcidlink}

\begin{document}


\title[Preprint submitted to XXX]{EEG Decoding for Datasets with Heterogenous Electrode Configurations using Transfer Learning Graph Neural Networks}
\author{Jinpei Han $^{1}$, Xiaoxi Wei $^{1}$, A. Aldo Faisal $^{1,2}$}

\address{%
$^{1}$ \quad  Brain \& Behaviour Lab, Department of Computing, Imperial College London, London SW7 2AZ, United Kingdom.\\
$^{2}$ \quad  Chair in Digital Health \& Data Science, University of Bayreuth, 95447 Bayreuth, Germany.}

\ead{aldo.faisal@imperial.ac.uk}
\vspace{12pt}
\begin{indented}
\item[]March 2023
\end{indented}

\begin{abstract}

\emph{Objective}. 
Brain-Machine Interfacing (BMI) has greatly benefited from adopting machine learning methods for feature learning that require extensive data for training, which are often unavailable from a single dataset. Yet, it is difficult to combine data across labs or even data within the same lab collected over the years due to the variation in recording equipment and electrode layouts resulting in shifts in data distribution, changes in data dimensionality, and altered identity of data dimensions. Our objective is to overcome this limitation and learn from many different and diverse datasets across labs with different experimental protocols. 
\emph{Approach}. 
To tackle the domain adaptation problem, we developed a novel machine learning framework combining graph neural networks (GNNs) and transfer learning methodologies for non-invasive Motor Imagery (MI) EEG decoding, as an example of BMI. Empirically, we focus on the challenges of learning from EEG data with different electrode layouts and varying numbers of electrodes. We utilise three MI EEG databases collected using very different numbers of EEG sensors (from 22 channels to 64) and layouts (from custom layouts to 10-20).
\emph{Main Results}. 
Our model achieved the highest accuracy with lower standard deviations on the testing datasets. This indicates that the GNN-based transfer learning framework can effectively aggregate knowledge from multiple datasets with different electrode layouts, leading to improved generalization in subject-independent MI EEG classification. 
\emph{Significance}. 
The findings of this study have important implications for Brain-Computer-Interface (BCI) research, as they highlight a promising method for overcoming the limitations posed by non-unified experimental setups. By enabling the integration of diverse datasets with varying electrode layouts, our proposed approach can help advance the development and application of BMI technologies.
\end{abstract}

\vspace{2pc}

%
\maketitle
%
%

\section{Introduction}
Integrating large disparate datasets collected over years and many different sites has led to rapid progress in large deep learning models in computer vision \cite{russakovsky2015imagenet} and natural language processing \cite{peng2019transfer}. However, in many bio-signal and medical data processing domains, data collection is not as standardized in terms of equipment, sensor arrangements, and protocols for data collection. A prime example is Electroencephalography (EEG), which use spans a broad range of medical and technological applications from sleep analysis \cite{aboalayon2016sleep, jia2020graphsleepnet} to non-invasive brain-computer interfaces (BCI). In the following, we focus on EEG decoding for non-invasive BCI \cite{rao2013brain} as a challenging example of machine learning.

In EEG-based BCI decoding, one of the most challenging areas is motor imagery (MI) decoding, where a user imagines certain actions, which trigger changes in brain signals. The computer decodes their intention from these changes in the signal in real-time \cite{rao2013brain,pfurtscheller1988mapping}.
Existing work based on MI EEG provides a non-invasive and portable BCI solution for the users to control external devices such as rehabilitation robotics, mind-controlled wheelchairs etc., by mental execution \cite{pfurtscheller2000brain,carlson2013brain,bensch2007nessi}. However, the MI EEG methods commonly suffer from data scarcity and heterogeneity of the training datasets, which result in poor generalization performance on new test subjects. Extensive research efforts have been devoted to developing machine learning algorithms to enable automatic MI classification from EEG signals.

\begin{figure*}[h]
    \includegraphics[width=\linewidth]{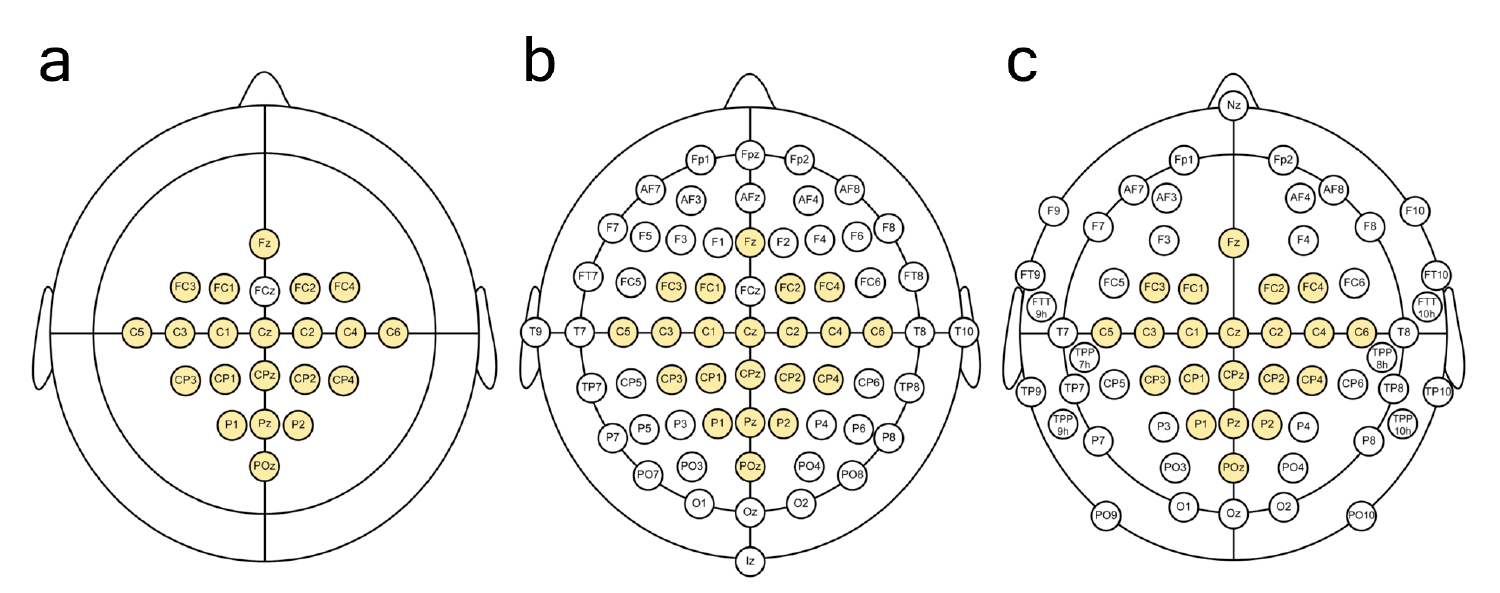}
    \caption{\textbf{The 3 motor imagery EEG datasets showing different numbers of electrodes and their arrangements, visualised in the EEG 10-20 system}: From left to right: BCIC 2a \cite{tangermann2012review}, PhysioNet MI \cite{schalk2004bci2000}, OpenBMI \cite{lee2019eeg} (from left).  A naive strategy to combine and utilise these datasets by only using the data recorded from these common sensors (yellow circles) discards a lot of data recorded from the other sensors (white circles).}
    \label{fig:figure1}
\end{figure*}

Deep learning advances have enabled end-to-end learning without prior feature extraction from raw EEG signals in recent years. Most existing MI-EEG studies employ deep learning methods based on two kinds of basic computational models, Convolutional Neural Network (CNN) \cite{walker2015deep,ortega2018compact,lawhern2018eegnet, schirrmeister2017deep, zhang2021eeg} and Recurrent Neural Network (RNN) \cite{wang2018lstm, zhang2018cascade}, that outperformed conventional Brain-Computer Interface (BCI) classifiers \cite{ang2008filter, subasi2010eeg}.  More recently, spatial and temporal attention mechanisms \cite{jia2020graphsleepnet,zhang2020motor} have also been introduced to EEG deep learning models to encourage the model to focus on more discriminative parts instead of processing the entire input equally. Feature learning requires deep learning models to use large datasets and delivers in turn the ability to learn intricate high-dimensional features, that often would have not been conceived by human-designed signal processing \cite{ortega2021hemcnn,ortega2021deep}.

Despite the success of deep learning methods in BCI decoding, and in most other domains of medical and biological signal analysis, there remain several issues concerning developing robust and accurate BCI decoding models using conventional end-to-end supervised training methods \cite{chen2022toward}. Deep learning methods are very data-hungry, and although efforts were made to improve their data efficiency in EEG BCI \cite{ferrante2015data,ortega2018compact,ponferrada2018data}, they still require a substantial amount of data. Moreover, data-efficient methods cannot address the heterogeneity issue caused by inter-subject variation. Inter-subject variations, in motor imagery BCI systems, are caused by different mental task conditions, and motivational or psychological states. Without a very large amount of training samples, existing deep learning methods are subject to overfitting which can lead to limited generalization performance on new subjects \cite{wei2021inter,chen2022toward,wu2020transfer}. However, since EEG signals are time-intensive to collect, the number of subjects and trials, even in publicly available datasets is relatively small \cite{ortega2020hygrip}. This is especially an issue if we are considering BCI interfaces for patients, where data collection burdens end-users and may impede the long-term adoption of technology \cite{makin2017neurocognitive}. In conclusion, the use of deep learning methods in BCI decoding has been successful. Still, it is data-intensive and requires a substantial amount of data to avoid overfitting caused by inter-subject variability. However, collecting a large amount of EEG data is time-intensive and can burden end-users, especially patients. To address these challenges, better and more principled methods are needed to improve the accuracy and robustness of BCI decoding models.

One of the most promising approaches to overcome these challenges is transfer learning. Transfer learning has been used successfully in many domains, including computer vision\cite{yu2022transfer}, and natural language processing\cite{ruder2019transfer}, to overcome the challenges of limited data and inter-domain variations by leveraging the knowledge learned from a related task \cite{wan2021review}. Increasing attention has been paid to addressing the challenging inter-subject variability and data scarcity issue using transfer learning \cite{zhang2022domain, li2019domain, han2020disentangled}, especially domain adaptation \cite{wan2021review}. We focus here on supervised domain adaptation as a special case of transfer learning to address the domain shift between different subjects and sensor layouts across multiple datasets. Supervised domain adaptation is a sub-field of transfer learning that aims to address the problem of training machine learning models in a source domain and adapting them to a target domain with different data distributions \cite{wan2021review}. This approach enables the transfer of knowledge from a well-labelled source domain to a poorly labelled target domain, thereby reducing data requirements and mitigating the problems of data scarcity and insufficient labelling as the model learned in a similar domain can be manually adjusted in the target domain with only a few trials \cite{wan2021review}.

%

The EEG Transfer Learning benchmark BEETL \cite{pmlr-v176-wei22a}, was initiated to evaluate the performance of current transfer learning algorithms across 5 motor imagery (MI) datasets with different numbers of channels, electrode locations and MI tasks. The top three winning teams used latent subject alignment, deep set alignment, label alignment, Maximum Classifier Discrepancy, or Riemannian geometry to achieve superior performance on the target dataset. While combining multiple datasets for training can facilitate the models' generalization performance, it is difficult to use data from multiple datasets for training since electrode layouts are often different. The collected data suffers from significant domain shifts resulting from a variety of devices, and experiment setups, which hindered the use of existing EEG deep learning models. While selecting common electrodes between datasets might be a practical solution, it may exclude information from the unique channels of that dataset \cite{pmlr-v176-wei22a}. The above solutions either used 17 common channels across all three source datasets to include more subjects or just used one source dataset to keep more common channels with the target dataset. The trade-off between keeping more common electrode channels or including more subjects from more datasets limited the effectiveness of transfer learning.

Unlike image data, where pixels are arranged in a grid, neural data has sensor arrangements better described by generic graphs. Existing CNN and RNN-based deep learning methods extract spatial features of the channels by stacking the sensor readings together in a grid, without taking into account the electrode positions \cite{demir2021eeg}, which neglects functional neural connections within the brain which oversimplifies the EEG feature extraction process \cite{demir2021eeg}. Thus, more sophisticated solutions are needed to deal more effectively with the variable sensor numbers, and their spatial relationships, which are seldom covered in the existing literature \cite{gu2023generalizable}. 

Inspired by the graph theory, Graph Neural Network (GNN) has increasingly attracted attention in deep learning research recently. It has the ability to capture complex relationships between objects and make inferences based on the graph structure data. Some studies have attempted to use GNN for EEG decoding to learn the neural connectivity of different parts of the brain \cite{song2018eeg,jia2020graphsleepnet,zhong2020eeg}. For example, Song \etal{} \cite{song2018eeg} and Jia \etal{} \cite{jia2020graphsleepnet} both proposed adjacency-matrix learning strategy to derive the intrinsic inter-channel relationships dynamically. However, their works need engineering some domain-specific features like differential entropy (DE) and power spectral density (PSD) from raw data and therefore are not end-to-end one-stage approaches. Zhong \etal{} \cite{zhong2020eeg} applied a regularized graph neural network architecture to learn the emotion-related functional connectivity with a node-wise domain adversarial training method to address the cross-subject variation during EEG-based emotion recognition. In addition, Demir \etal{} \cite{demir2021eeg,demir2022eeg} proposed EEG-GNN and EEG-GAT to classify Errp and RSVP EEG signals. The latter EEG-GAT model involves a multi-head attention mechanism to enhance the functional neural connectivity subject to specific cognitive tasks between different electrode sites. However, empirical results showed GNNs are prone to suffer from overfitting and poor convergence for small datasets, the issue of data scarcity and heterogeneity in MI EEG datasets might be more prominent \cite{demir2021eeg}. Special architecture designs are needed for GNN models to adapt to EEG data from unseen subjects or datasets.

In this work, we propose a framework combining GNNs with transfer learning that can aggregate EEG data with different electrode layouts and facilitate data generalisation from the target dataset. Our framework consists of separate GNN blocks to learn the spatial connectivity of each individual EEG dataset with different electrode layouts, followed by a shared latent alignment block to handle both subject-level and dataset-level heterogeneity. Our proposed framework can effectively be trained from multiple datasets simultaneously. To compare with the experimental settings in existing GNN research and transfer learning studies for EEG signals, here we highlight the novelty of our learning paradigm. First, conventional GNN models for EEG decoding are validated under subject-dependent settings. In this study, the training and test sets are different groups of subjects thus making the evaluation of the algorithm subject-independent, which test the generality of the algorithm. Secondly, existing works on transfer learning in the EEG domain mostly focus on inter-subject adaptation within the same database. A portion of the target subjects' EEG data is generally available for training in the experiment setting of supervised transfer learning works, and with both the data from source subjects and target subjects for training, the model can be quickly and accurately adapted. In contrast, our work focuses on utilising the data from multiple small EEG datasets with different sensor layouts and making our GNN model adapt to EEG data of fully unseen subjects in the target dataset. We present an early work that uses GNN and domain adaptation to simultaneously address the inter-subject variation and the cross-dataset channel variations. 

\section{Materials and Methods}
\begin{figure*}[htbp]
\centering
\includegraphics[width=\linewidth]{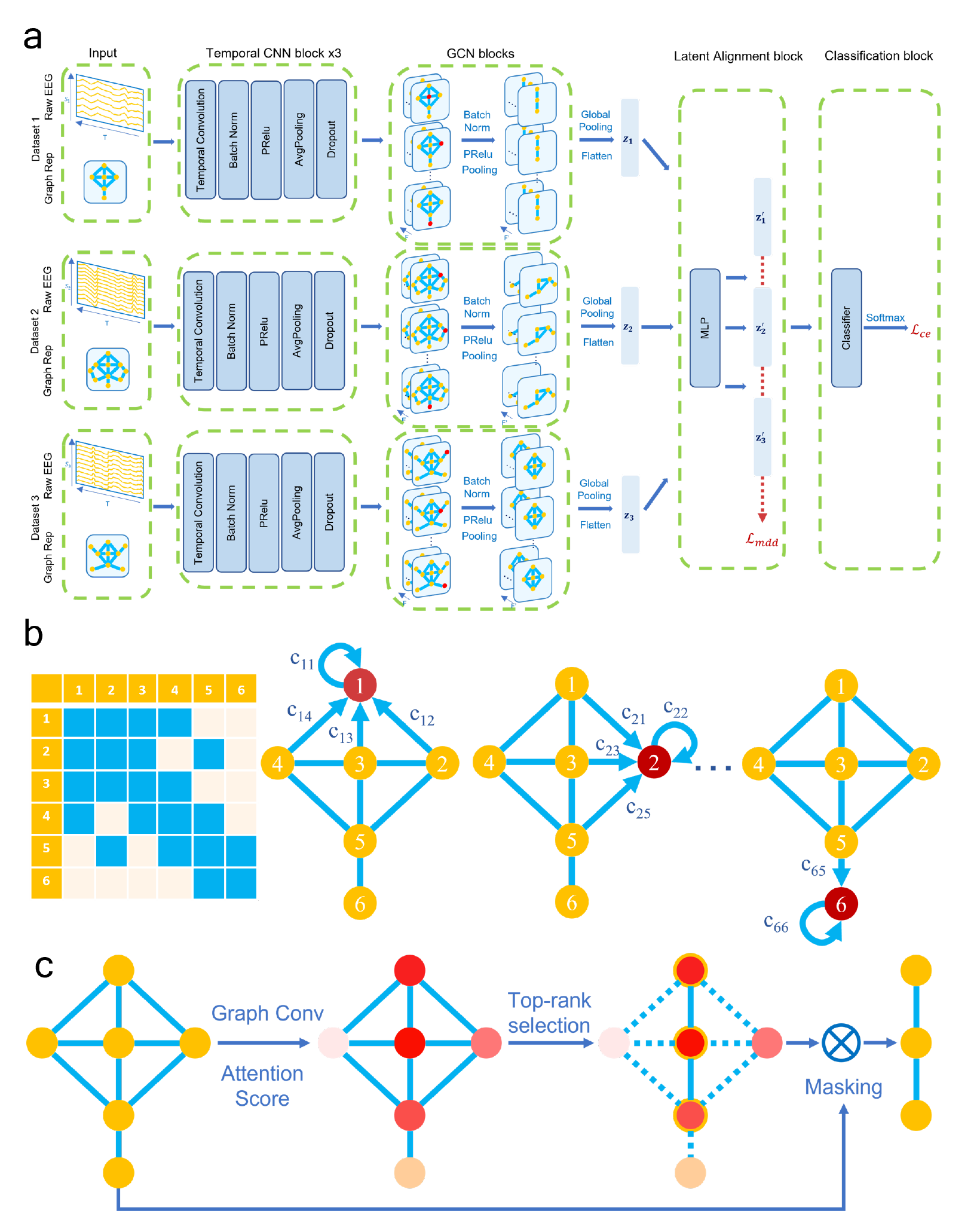}
\caption{\textbf{(a) Our proposed frameworks for training with multiple motor imagery EEG datasets.} The framework starts with Temporal CNN blocks with similar architecture to EEGNet \cite{lawhern2018eegnet} without spatial filters. After each temporal convolution layer, we apply batch normalization, PRelu activation layer, average pooling and dropout layers to introduce non-linearity and reduce the risk of overfitting. After each GCN block, the number of channels is reduced by approximately a factor of 2. At the final stage of the GCN block, we apply the global mean pooling operator, which averages the remaining node features and produces a graph feature vector. The feature vector passes through an MLP latent alignment block and finally, a Linear classifier was used to perform the classification. \textbf{(b) Illustration of the adjacency matrix and the local aggregation process in each GCN block.} \textbf{(c) Illustration of the SAGPool process in each GCN block.}}
\label{fig:figure2}
\end{figure*}

To tackle the real-world challenges, we apply supervised domain adaptation to a GNN framework to aggregate information from multiple small datasets with different electrode layouts. The overall pipeline of the proposed framework is summarised in Figure \ref{fig:figure2}(a). The proposed framework consists of separate encoders $E_d, G_d$ for each dataset, a shared latent alignment layer $P$ and a shared classifier $C$. Each encoder branch inputs the fixed dimensionality of the EEG signal to the particular dataset. The shared latent alignment block $P$ uses distance metric to simultaneously address the distribution shifts across different subjects and different electrode layouts. During inference time, only the corresponding encoder branch is used to decode the data from the target dataset.
\subsection{Problem Formulation}
To formulate the problem in mathematical terms, consider EEG segments from $D$ different datasets. We denote the input into our neural networks from dataset $d \in D$ as $\mathbf{X}\in\mathbb{R}^{S_d\times T}$, where $S_d$ represent the number of sensors of EEG recordings from dataset $d$ and $T$ represent EEG samples with size of recording frequency times window size in seconds. The MI class output is represented as $y \in [0, 1]$, where class 0,1 represents left/right-hand imagination, respectively. An adjacency matrix that defines the graph structure and the edge weights of EEG recordings from dataset $d$, is represented as $\mathbf{A}_d=S_d \times S_d$, where an undirected edge between two sensors $i,j$ is represented as $(\mathbf{A}_{ij}>0)$. In this work, we focus on utilising the data from multiple small EEG datasets to form different graph representations and use transfer learning to make our GNN model adapt to EEG data with different sensor layouts.


\subsection{Graph Neural Network Background}
To apply our proposed framework, we utilize GNNs to process the EEG data represented as graphs.  A GNN is a type of artificial neural network that can process data that is represented as graphs. It is widely used to learn the relationships between elements in a graph structure, such as in knowledge graphs, social networks, and chemical compounds. 

Forming graph-structured data for a GNN typically involves representing the graph as a combination of two main components: the graph structure and the node features. The graph structure is typically represented using an adjacency matrix or an edge list. An adjacency matrix is a square matrix of size $S \times S$, where $S$ is the number of nodes in the graph, and the entry $\mathbf{A}_{ij} > 0$ if there is an edge between nodes $i$ and $j$, and 0 otherwise, as shown in Figure \ref{fig:figure2}(b). An edge list is a list of tuples $(i, j)$ representing the edges in the graph, where $i$ and $j$ are the indices of the nodes that are connected by the edge. The node features are typically represented as a feature matrix $\mathbf{X} \in \mathbb{R}^{S\times F}$, where $S$ is the number of nodes in the graph and $F$ is the number of features for each node. Each row of the feature matrix corresponds to a node in the graph and the columns represent the features associated with that node.
 Once the graph structure and node features are represented, they can be used as input to a GNN. A GNN uses the graph structure to propagate information across the graph and the node features to perform computations at each node. GNN is composed of several layers, each layer having its own set of parameters. The layers are connected in a graph structure, where each node in the graph corresponds to a certain parameter of the layer, and the edges between nodes represent the connections between the parameters. At each layer, the GNN takes in the data from the previous layer and applies a function to it to produce the output for the current layer. This process is called message passing or graph convolution. Graph Convolutional Neural Network (GCN) \cite{welling2016semi} is the most commonly used variant of GNN in the literature. In GCNs,  the adjacency matrix $\mathbf{A}$ is normalized by $\mathbf{\hat{A}} = \mathbf{\tilde{D}}^{-1/2} \mathbf{\tilde{A}} \mathbf{\tilde{D}}^{-1/2}$, where self-connections are enforced by $\mathbf{\tilde{A}} = \mathbf{A} + \mathbf{I}$. $\mathbf{\tilde{D}}_{ij}$ is the diagonal node degree matrix of the graph, which performs a row-wise summation of the adjacency matrix $\mathbf{\tilde{D}}_{ij} = \sum_j \mathbf{\tilde{A}}_{ij}$, which is the degree of each node. The graph convolution operation allows for the incorporation of the structural information of the graph into the convolution operation. The graph convolution operation is defined as follows:
\begin{equation}
    \mathbf{X}^{(l+1)} = \sigma(\mathbf{\hat{A}} \mathbf{X}^{(l)}  \mathbf{\Theta}^{(l)})
\end{equation}

Where $\mathbf{\hat{A}}$ is the symmetric normalised adjacency matrix, $\mathbf{X}^{(l)}$ is the node feature matrix at layer $l$, and $ \mathbf{\Theta}^{(l)}$ is the trainable weight matrix at layer $l$, $\sigma$ is a non-linear activation function, and $\mathbf{X}^{(l+1)}$ is the output feature matrix at layer $l+1$. from a node-wise local aggregation perspective, graph convolution is performed by aggregating information from the neighbourhood of a node, and then updating the representation of that node based on the aggregated information iteratively: 
\begin{equation}
     x_i^{(l+1)} = \sigma(\mathbf{\Theta}^{\top} \sum_{k \in \mathcal{N}(i) \cup \{ i \}} c_{ij} x_j^{(l)})
\end{equation}
Where $x_i^{(l+1)}$ is the representation of node $i$ at layer $l+1$, $ \mathcal{N}(i)$ is the set of neighbours of node $i$, and $c_{ij}$ is the weight associated with the edge between node $i$ and node $j$. The process is visualised in Figure \ref{fig:figure2}(b). In this way, the graph convolution operation iteratively updates the feature representation at each node, taking into account the adjacency matrix of the graph in order to propagate information across the graph and effectively learn patterns in the graph structure.


Graph coarsening and pooling layers are important for GNN models to avoid overfitting by reducing the number of parameters \cite{lee2019self}. Self-Attention Graph Pooling (SAGPooling) \cite{lee2019self, knyazev2019understanding} aims to aggregate the information of a graph's nodes and reduce the graph's size for further processing. SAGPooling is a type of graph pooling which is based on self-attention mechanisms. It can be used to reduce the computation and memory requirements as well as improve the performance of GNNs due to its ability to capture the graph's global structure, as well as the relationships between the nodes. The main idea behind SAGPooling is to use a self-attention mechanism to weigh the importance of each node in the graph, and then pool the nodes by keeping only the most important ones. The self-attention mechanism is used to compute the attention scores for each node, which represents the importance of that node in the graph. The attention scores $\mathbf{Z}$ are computed using a GNN layer by 
\begin{equation}
    \mathbf{Z}^{(l)} = (\mbox{GNN}(\mathbf{H}^{(l)}, \mathbf{\hat{A}}^{(l)})
\end{equation}
The pooling process is then performed by keeping only the top-k nodes with the highest attention scores, $\mbox{idx} = \mbox{top}_{\mbox{k}}(Z_{\mbox{att}}^{(l)})$, and discarding the rest. Where $\mbox{idx}$ is the index of the new set of selected nodes and $\mbox{top}_{\mbox{k}}$ is the function that selects the top k nodes based on their attention scores as shown in Figure \ref{fig:figure2}(c). The remaining nodes' feature matrix and adjacency matrix are updated with $\mathbf{X}^{(l+1)} = \mathbf{X}^{(l)} \odot \mathbf{Z}_{\mbox{idx}}$ and $\mathbf{\hat{A}}^{(l+1)} = \mathbf{\hat{A}}^{(l)}_{\mbox{idx}, \mbox{idx}}$, where $\odot$ is the elementwise product \cite{lee2019self}.  

\subsection{Representing EEGs as Graphs}
\begin{figure*}[htbp]
    \centering
    \includegraphics[width=\linewidth]{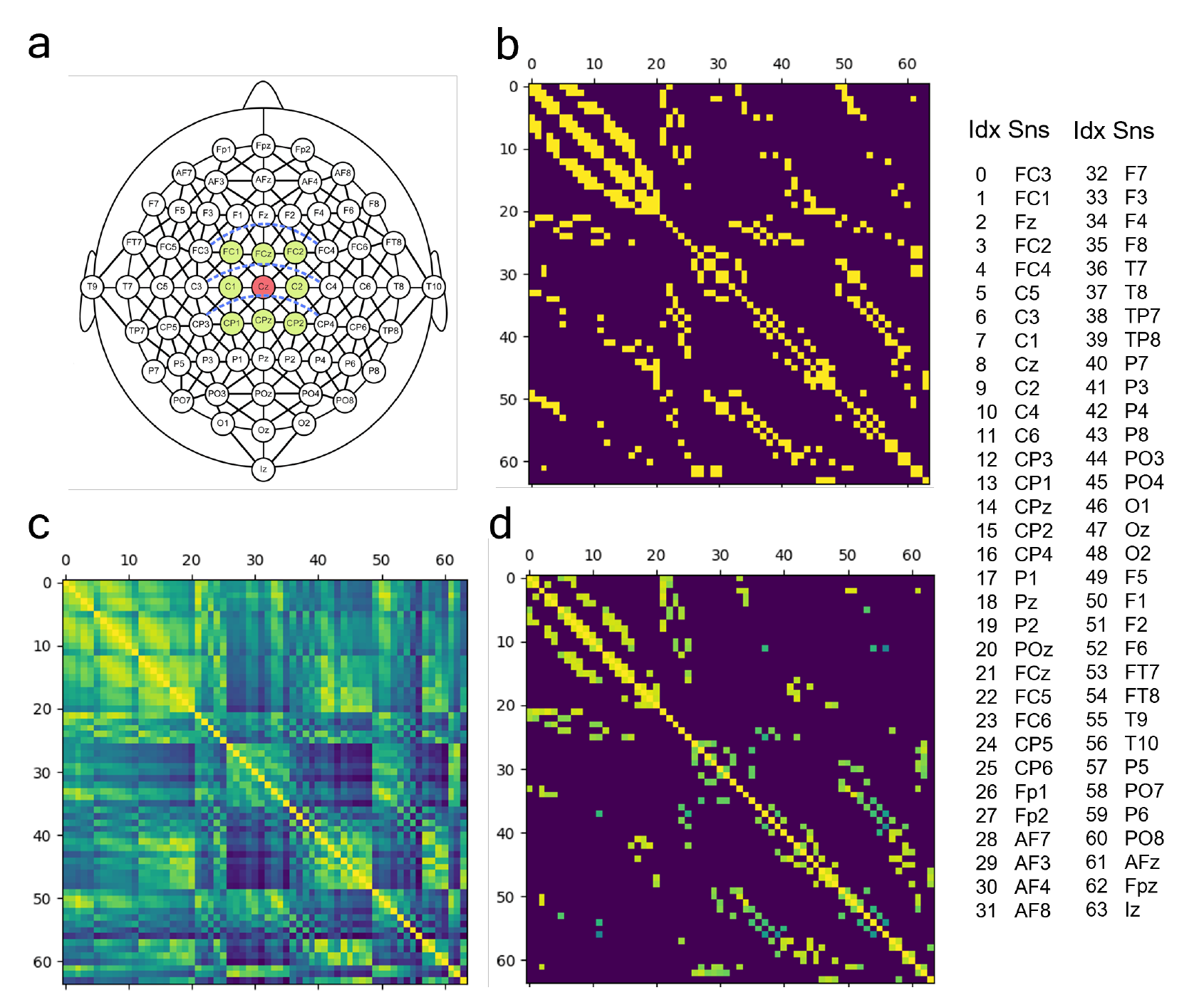}
    \caption{\textbf{(a) Illustration of the PhysioNet MI sensor grid representation on the EEG 10-20 system and (b) its adjacency matrix $\mathbf{A}$ with shape $64 \times 64$, constructed using the neighbourhood method}. We draw here the full graph with sensors and their immediate neighbours connected by a black line. As an illustration of the neighbourhood concept, $C_z$ (red node) is connected to its immediate neighbour nodes (green nodes). Additional global connections are shown in blue. Note, that the colour coding is binary in the adjacency matrix (right), as we only consider connections that exist (yellow) and that do not exist (dark blue). \textbf{ (c) The adjacency matrix constructed using the correlation method. (b) The resulting adjacency matrix when only the top 5 connections for each node are kept.} The numerical index of each sensor is shown on the right. }
    \label{fig:figure3}
\end{figure*}

To process EEG data using GNN, We first need to recast the EEG sensor grid as a graph, whose node features are the sensor time-series of EEG data (see Figure~\ref{fig:figure3}(a)). The process of representing EEGs as graphs involves defining an adjacency matrix that captures the relationships between the different EEG sensors. This matrix is critical to the process of graph representation learning, which is a fundamental part of using GNNs to analyze EEG data. We define an adjacency matrix $\mathbf{A} \in \mathbb{R}^{S_d\times S_d}$ of the EEG sensors and is essential to graph representation learning. In this work, we used two different methods to form the adjacency matrix, namely the neighbourhood method and the correlation method. 

The neighbourhood method uses the physical geometry of the EEG sensor grid. We define the graph connections using natural EEG electrode layouts as shown in Figure \ref{fig:figure3}(b), where each node is connected to its direct natural neighbours \cite{zhang2020motor}, which are the surrounding electrodes of the node on the scalp. Prior studies has demonstrated the significance of asymmetry in brain activity between the left and right motor cortex for MI prediction \cite{zhong2020eeg}. To propagate information across both sides, numerous graph convolution layers are required, as each layer primarily aggregates data from a node's immediate neighbors. However, this approach can result in over-squashing issues in GNNs, where the node representations lose essential information and become indistinguishable due to excessive compression through multiple layers of non-linear transformations when aggregating messages over a long path, ultimately diminishing the network's capacity to represent the graph's intricate structure \cite{alon2020bottleneck}. Therefore, we added additional global connections between (FC3, FC4), (C3, C4), and (CP3, CP4). These nodes are shared across all three datasets and have large numbers of neighbour nodes within the motor cortex, which capture the EEG asymmetry during motor imagery tasks. In this way, the graph formation method can simultaneously represent the local electrode connectivity and MI-related global functional connectivity of different electrode layouts. The adjacency matrix forms a spatial-temporal graph with the EEG data $\mathbf{X}$ where each node feature is the EEG sensor recording values $T$.

Our second method is the correlation method, which uses the functional correlation between EEG signals from different sensors to determine the neighbourhood relationships. We used the Absolute Pearson Correlation Coefficient method to capture the dynamic connectivity of different parts of the brain \cite{tang2021self}. We initialised a fully connected graph structure for the EEG sensors, where the weight of each edge $\mathbf{A_{ij}}$ is the absolute value of the normalised cross-correlation between sensor $i$ and sensor $j$. We only kept the edges with top $k$ weight values and remove the rest to introduce sparsity and reduce computation cost. The adjacency matrix constructed by the PCC method averaged across all subjects in the PhysioNet dataset is shown in Figure~\ref{fig:figure3}(c), where we used the 5 largest correlation values belonging to a given node, to define the neighbourhood as shown in Figure~\ref{fig:figure3}(d).

\subsection{Temporal/Spatial Feature Learning}
We first start our decoding process by extracting the temporal features from the EEG signals. The input into our neural networks from each dataset is a tuple of data $\mathbf{X}=S\times T$ of values from $S$ EEG sensors times $T$ samples and second, a normalised adjacency matrix $\mathbf{A}=S \times S$. Empirical results reported from several EEG studies \cite{demir2021eeg, demir2022eeg, wu2020comprehensive} suggested that GNN models are prone to overfitting as the feature dimension for each node increases. For MI EEG data, the node feature for each electrode would be the recordings in the time domain $T$, which usually contains hundreds of data points for each trial. Thus it is essential to perform feature extraction on the EEG data prior to the graph convolution. Although there have been studies \cite{song2018eeg} that used conventional feature extraction methods, such as the frequency band features, we decided to use CNN to extract temporal features in an end-to-end setting. We used the temporal CNN block with large kernel sizes and pooling layers on the time axis to extract temporal EEG features and dramatically lower the node features dimensions to $F$. Each CNN block consists of a 1-D convolution layer followed by batch normalisation, PReLU activation, average pooling and drop-out layers to prevent overfitting. Since the PReLU activation function achieves better performance than the ReLU in a shallow encoder network \cite{he2015delving}.

The node feature matrix $\mathbf{X}$ with size $S \times F$ and the normalized adjacency matrix $\mathbf{\tilde{A}}$ are passed to the GNN block $G$ for spatial feature learning. Each GCN layer performs graph convolution and produces the output $\mathbf{X^{\prime}}$. After each graph convolution layer, we employ batch normalization and the SAGPooling strategy \cite{lee2019self} to learn the top $k$ most important nodes and drop the rest of the nodes. In the final layer of our GNN block, We used global mean pooling which averages the features across the remaining nodes and flattens the graph feature as a latent vector $z$. The latent vectors of each decoding branch are concatenated in the batch dimension and fed into the latent alignment block. We presented our method to learn the temporal and spatial features from EEG signals using a combination of CNNs and GNNs. By leveraging the advantages of both techniques, we were able to efficiently extract relevant features and reduce overfitting, ultimately resulting in a robust and end-to-end EEG decoding framework.

\subsection{Latent Feature Alignment}
Building upon the feature extraction and graph neural network framework established in the previous subsection, we now introduce the latent alignment block, which is designed to perform domain adaptation and ensure the generalizability of our model across various datasets. The latent feature alignment block consists of a shared MLP projector $P$ that projects latent vector $z$ of different datasets to $z'$ and aligns their distribution in latent space. We used Maximum Density Divergence (MDD) losses \cite{li2020maximum} to simultaneously minimize the inter-dataset divergence and maximize the intra-MI class density. The MDD loss can encourage the dataset-specific encoders to extract dataset-invariant, task-related features. 
Given the set of latent features $z_1$ from 1 and $z_2$ from dataset 2, we calculate the pair-wise MDD loss by:

\begin{equation}
\small
\begin{aligned}
    \mathcal{L}_{mdd} &= \frac{1}{N}\sum^{N}_{i} \| P(z_{1, i}) - P(z_{2, i}) \| ^2_2 \\
    &+ \frac{1}{M_{1}} \sum_{y_{1, i} = y_{1, j}} \| P(z_{1, i}) - P(z_{1, j}) \| ^2_2 \\
    &+ \frac{1}{M_{2}} \sum_{y_{2, i} = y_{2, j}} \| P(z_{2, i}) - P(z_{2, j}) \| ^2_2   
\end{aligned}
    \label{equ:mdd}
\end{equation}
where, $N$ is the batch size for each dataset, $M_1$ and $M_2$ are the numbers of samples with the same labels in a batch. Where the first term is the sum of the mean distance between the target and source features, the second term is the mean distance between features with the same class labels within the target dataset and the third term is the mean distance between features with the same class labels within the source dataset. The total MDD loss, $\mathcal{L}_{\mbox{mdd}}$ is the sum of the pair-wise MDD loss between the feature of two datasets: $\mathcal{L}_{\mbox{mdd}} = \sum^{N}_{i,j}\mbox{MDD}(z'_i, z'_j) (i \neq j)$, where $N$ is the number of different datasets and $i, j$ represents the different index of datasets. Since there are different subjects in different datasets, the MDD loss can simultaneously minimise the inter-subject distribution shift of latent features. We finally calculate the cross-entropy loss $\mathcal{L}_{\mbox{ce}}$ from the outputs of a Linear classification layer $C$ and the supervised labels.

Overall all blocks and the classifier are updated based on $w_1\mathcal{L}_{\mbox{ce}}+ w_2\mathcal{L}_{\mbox{mdd}}$, where $w_1, w_2$ are the weights for each part of the loss. At the inference stage, the raw EEG sequence passes through the target temporal CNN block, GNN block, the shared latent alignment block and the classifier to predict the imagined actions. 

\subsection{Experimental Settings}
We picked 3 publicly available Motor Imagery (MI) datasets with various numbers of channels and electrode layouts. The size of each dataset was limited to simulate the many small-scale EEG datasets available and balance the total number of trials from each dataset. 
\begin{itemize}
    \item The BCIC IV 2a dataset \cite{tangermann2012review} consists of 22 channels of EEG data from 9 different subjects, with each subject performing 144 trials of imagining left hand, right hand, feet, and tongue movements.
    \item The PhysioNet MI dataset \cite{schalk2004bci2000} comprises 64-channel EEG data recorded from 109 different subjects. Each subject was asked to imagine opening and closing their left and right fist in 45 trials. We used the first four seconds of each trial and included only the first 57 subjects in our analysis.
    \item The OpenBMI dataset features 62-channel EEG signals recorded from 54 different subjects. Each subject was asked to imagine grasping their right or left hand for 4 seconds in 400 trials. We selected only the first 18 subjects for our study since this dataset has a significantly larger number of trials per subject
\end{itemize}

For our experiment, we included the right-hand and left-hand MI tasks from each dataset and used Inter-Subject Cross-Validation on each target dataset. We treated one of the three datasets as the target dataset while the other two as source datasets in each experiment. The data of some subjects in the target dataset is made available as supporting examples for transfer learning, while the remaining subjects' data were used for testing in an inter-subject cross-validation setting. We included different numbers of subjects from different datasets for the experiments since the number of trials per subject is different. In each fold, we try to use a similar amount of training and testing data trials for each dataset as shown in Table \ref{tab:experiment}. We used the first $4$s of each trial across all the datasets, bandpass filtered between $4-40$Hz, downsampled to $125$Hz, and normalized using the z-score normalization method. We used the average Accuracy and F1 score were used as the performance metrics.

    

\begin{table*}[htbp]
  \centering
  \setlength{\tabcolsep}{3pt}
  \caption{Description of Datasets Used for Experiments.}
    \begin{tabular}{cccccc}
    \toprule
    \multirow{2}[0]{*}{\textbf{Dataset}} & \multirow{2}[0]{*}{\textbf{\#Channel}} & \multicolumn{2}{c}{\textbf{Train}} & \multicolumn{2}{c}{\textbf{Test}} \\
    &       & \#Subject & \#Trial & \#Subject & \#Trial \\ \midrule
    BCIC 2a & 22    & 8     & 2304  & 1     & 288 \\
    PhysioNet & 64    & 51    & 2295  & 6     & 270 \\
    OpenBMI & 62    & 11    & 2200  & 1     & 200 \\ \bottomrule
    \end{tabular}%
  \label{tab:experiment}%
\end{table*}%
In order to address the trade-off between using more datasets or more channels, we consider three baseline setups that highlight the different approaches. The primary challenge arises from the varying number of channels and channel arrangements present in the three datasets, which can impact the performance of traditional deep learning models. Our baseline setups are as follows:
\begin{itemize}
    \item Single CNN 1: A CNN trained using only the supporting examples in the target dataset, focusing on maximizing the use of available channels within a single dataset.
    \item Single CNN 2: A CNN trained using common channels across all three datasets, which aims to leverage the combined information from multiple datasets but at the cost of utilizing fewer channels.
    \item Triple CNN: A model with three separate CNN encoders, each employing different spatial convolution kernels in place of GNN encoders. This approach incorporates latent alignment using MDD before the shared classification layer, resembling the overall structure of SCSN \cite{wei2021inter}. By doing so, this baseline setup can leverage multiple datasets while neglecting the different channel arrangements.
\end{itemize}

For our proposed method, we tested our framework with the neighbourhood adjacency matrix (Proposed + $\mathbf{A}_n$) and with the correlation adjacency matrix (Proposed + $\mathbf{A}_c$).

We used CNN encoders with a similar structure to the EEGNet \cite{lawhern2018eegnet} for our baseline methods while splitting it into a temporal feature extraction block and a spatial feature extraction block same as our proposed framework. We implemented all models using open-source software \emph{Python}, \emph{Braindecode}, \emph{PyTorch} and \emph{PyTorch-Geometric}. We used the Adam optimizer (learning rate of $0.005$, beta $\in [0.9, 0.999]$, weight decay of $0.0001$) and batch size for each dataset of $256$  and trained on a single NVIDIA RTX-3090 GPU powered Linux workstation.  
\section{Results}
\begin{figure*}[htbp]
    \centering
    \includegraphics[width=\linewidth]{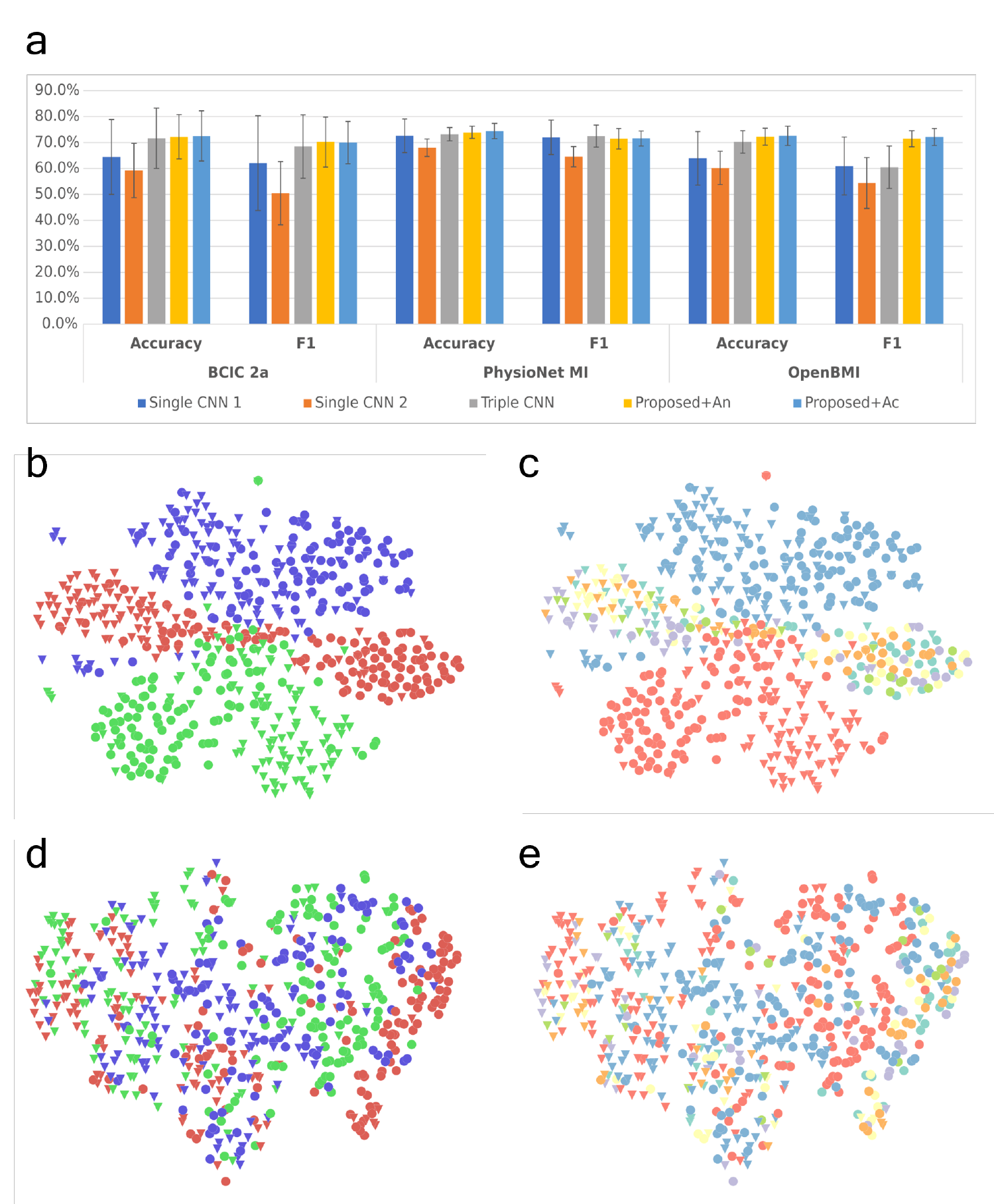}
    \caption{\textbf{(a) Average inter-subject cross-validation results.} Single CNN 1: A single CNN model trained with the training samples from a single dataset, Single CNN 2: A single CNN model trained with samples from all datasets by selecting the common channels. \textbf{tSNE plots of the latent features of test data from three EEG datasets without latent alignment:} \textbf{(b)} Colours represent different datasets. \textbf{(c)} Colours represent different subjects, Circle and Square represent left and right motor imagery classes respectively. \textbf{tSNE plots of the latent features of test data from three EEG datasets with latent alignment:} \textbf{(d)} Colours represent different datasets. \textbf{(e)} Colours represent different subjects, Circle and Square represent left and right motor imagery classes respectively.}
    \label{fig:figure4}
\end{figure*}

The cross-subject group level classification results are presented in Table~\ref{result} and Figure~\ref{fig:figure4}(a). Our method achieved the highest average performance scores compared to the 3 baseline methods. Specifically, the correlation graph method achieved 72.5\%, 74.4\%, and 72.6\% accuracies on the BCIC 2a, PhysioNet MI, and the OpenBMI dataset, respectively. The standard deviation of the accuracy and F1 score across different folds achieved by our methods are smaller compared to other methods, which indicates that our methods are more stable and robust against the subject variations in the test dataset. 


The Single CNN 1 model trained using only the target dataset performed poorly on the BCIC 2a and OpenBMI datasets with accuracies of 64.5\% and 63.9\%, respectively. The poor performance is likely due caused by overfitting due to its limitation on data ingestion, especially for the BCIC dataset, which only includes data from 8 subjects for training in each fold. The Single CNN 1 model performed better for the PhysioNet dataset since it includes more numbers of subjects as examples for training.

The Single CNN 2 that directly trained with data from multiple datasets by using all channels they had in common, without any alignment performed worse. It achieved 5.30\% lower accuracy than the Single CNN1 setup on the BCIC dataset with only 21 common channels. This is likely due to using only the subset of common electrodes across datasets leading to the loss of information. Moreover, heterogeneity across datasets caused by different electrode layouts, and negative transfer across different subjects will have decreased the performance \cite{wei2021inter}. Thus, selecting only the common shared channels between datasets to combine them is truly limiting. 

The Triple CNN model, which incorporates separate CNN blocks and a common classifier with latent alignment, significantly outperforms the Single CNN models on both the BCIC 2a (ACC: 71.6\%) and OpenBMI datasets (ACC: 70.2\%). This improvement in performance confirms the effectiveness of our latent alignment strategies, as the Triple CNN model is able to successfully leverage information from multiple datasets and achieve superior results compared to the traditional Single CNN methods. However, it achieved lower performance than both of our proposed GNN models due to the spatial convolution filtering without taking into account the spatial relationships between the sensors.

\begin{table*}[htbp]
  \centering
  \caption{Average Accuracy and F1 Score of Inter-subject Cross-Validation.}
   \resizebox{\linewidth}{!}{
    \begin{tabular}{ccccccccc}
    \toprule
    \multirow{2}[0]{*}{\textbf{Method}} & \multirow{2}[0]{*}{\textbf{Channel}} & \multirow{2}[0]{*}{\textbf{Train Set}} & \multicolumn{2}{c}{\textbf{BCIC 2a}} & \multicolumn{2}{c}{\textbf{PhysioNet MI}} & \multicolumn{2}{c}{\textbf{OpenBMI}} \\
          &       &       & Accuracy & F1    & Accuracy & F1    & Accuracy & F1 \\
    \midrule
    Single CNN 1 & All   & Single & 64.5\%   &  62.1\%   &   72.6\%    &   72.0\%   & 63.9\% & 61.0\% \\
    Single CNN 2 & Common & All   &    59.2\%   &   50.5\%    &   68.0\%    &   64.6\%   & 60.2\% & 54.4\% \\
    Triple CNN & All   & All   &    71.6\%   &   68.5\%    &    73.2\%   &   72.5\%    & 70.2\% & 60.5\% \\
    Proposed + $\mathbf{A}_n$ & All & All& 72.2\%  &70.2\%  & 73.9\% & 71.5\%  &72.3\% & 71.5\% \\
    Proposed + $\mathbf{A}_c$ & All & All&  72.5\%  & 70.0\%  & 74.4\%  & 71.6\%  &72.6\% & 72.2\% \\
    \bottomrule
    \end{tabular}}%
  \label{result}%
\end{table*}%

\subsection{Interpreting what the GNN  learned}

We next investigated the internal represetnations of the GNN. To this end we use t-SNE to visualise the effect of our latent alignment method in the feature space. Figure~\ref{fig:figure4}(b) and (c) show the t-SNE projection of the MI embedding features of test data produced by the Single CNN 2 model, which does not have a latent alignment block. In Figure~\ref{fig:figure4}(b), the red, green and blue points represent the latent features extracted from the three different datasets respectively. The two different shapes, circle and square represent the left and right MI classes. The latent features of the same colour (same dataset) clustered together in the tSNE plot, which indicate that the extracted features contain a lot of dataset-specific information. Moreover, Figure~\ref{fig:figure4}(c) uses various colours to represent the latent features extracted from different subjects. The clustering of the same-coloured latent features (from the same subject) indicates the presence of subject-specific information in the extracted latent features. Consequently, without the latent alignment block to address domain shifts across subjects and datasets, a clear decision boundary is not apparent in the learned latent space.

In contrast, Figure~\ref{fig:figure4}(d) and (e) present the MI embedding features generated by our proposed method. It is evident that the majority of the latent features cluster based on their shapes (MI classes) rather than their colours (datasets/subjects). Specifically, circles cluster on the right, and triangles cluster on the left. This result demonstrates that our domain adaptation method enabled the MI encoder to effectively learn MI features from all three datasets while suppressing dataset/subject-specific features.

This look into the GNN-black box allow us to inform our interpretation why our proposed methods outperforms the 3 reference approaches.


\subsection{Ablation Study}
We further conducted ablation studies to validate the effectiveness of both the GNN encoder and the latent alignment module of our proposed method, with the results shown in Table \ref{tab:ablation}. All ablation settings used all the channels from the three datasets. The first ablation setup used a triple CNN encoder with a shared classifier but without latent alignment. The second ablation setup added GNN for spatial feature extraction in addition to the CNN encoder. Both of these methods failed to achieve optimal inter-subject classification performance, as they lacked specific mechanisms to handle the subject variations. The third ablation setup is the same as the Triple CNN setup in the baseline experiment, which is the first ablation setup with the addition of the latent alignment mechanism. Although this setup achieved higher classification performance than the previous two configurations due to the presence of the latent alignment layer, our proposed framework with the GNN encoder and latent alignment proved to be the most effective. These ablation studies demonstrate the importance of both the GNN encoder and the latent alignment module in our proposed method, as they contribute to the optimal inter-subject classification performance across different datasets.
\begin{table*}[tbp!]
\centering
  \caption{Average classification Accuracy and F1 score based on different model ablation designs}
  \label{tab:ablation}%
      \begin{tabular}{cccccccc}
        \toprule
        \multicolumn{2}{c} {\textbf{Ablation}} & \multicolumn{2}{c}{\textbf{BCIC 2a}} & \multicolumn{2}{c}{\textbf{PhysioNet MI}} & \multicolumn{2}{c}{\textbf{OpenBMI}}\\
         $\boldsymbol{GNN}$  & $\boldsymbol{L}_{mdd}$  & Acc (\%) & F1 & Acc (\%) & F1 & Acc (\%) & F1\\ \midrule
          ~ &  ~ & 66.2\% & 64.3\% & 71.9\% & 70.6\% & 67.9\% & 60.3\% \\
         \checkmark &  ~ & 68.9\% & 69.1\% & 71.7\% & 69.8\% & 69.1\% & 62.5\% \\
          ~ & \checkmark & 71.6\% & 68.5\% & 73.2\% & 72.5\% & 70.2\% & 68.5\% \\
         \checkmark & \checkmark  & 72.5\% & 70.0\% & 74.4\% & 71.6\% & 72.6\% & 72.2\% \\
         \bottomrule
      \end{tabular}
\end{table*}

\section{Discussion}

Our proposed GNN-transfer learning framework achieved the best performance validated using three different datasets and outperforms the popular CNN methods. The results demonstrated that we can use GNNs to aggregate EEG data from multiple very different datasets with heterogeneous electrode configurations. Our GNN framework extracts spatiotemporal features of EEG data. It performs domain adaptation using MDD loss to align the feature distributions in latent space, which enforces the framework to learn features that can generalise across different subjects/datasets. This required us to translate these techniques and map them onto GNNs by first representing EEG data as graph representations using adjacency matrices, then extracting temporal features using CNNs as node features, and using GNN to learn the spatial connection of different electrode configurations.

Compared to other GNN studies in the EEG literature, most of the existing studies used handcrafted features extracted from different EEG bandwidths \cite{song2018eeg, jia2020graphsleepnet}, with a few other end-to-end methods that extract temporal features in the decoding process\cite{demir2021eeg, demir2022eeg}. However, those existing decoders were only learned on a single dataset during experiments and GNNs are found to be prone to overfitting \cite{demir2021eeg, qiu2019rethinking, wu2019session}. In our study, we introduced transfer learning methods, specifically domain adaptation methods to circumvent the overfitting issue of GNN and trained our framework using data from multiple datasets.

Although domain adaptation has been widely used in the literature to address the inter-subject or inter-session variations in MI EEG data \cite{hang2019cross, zhao2020deep, tang2020conditional}, it is much less explored in the literature when it comes to handling dimensionality changes in input features  \cite{gu2023generalizable}. Specifically for the EEG data, this refers to scenarios where data was collected using different sensing modalities and spatial-temporal resolutions \cite{gao2019hhhfl, ju2020federated}. Gao \etal{} \cite{gao2019hhhfl} studied EEG decoding with the signals collected from multiple devices with different numbers of channels. The authors projected the EEG recordings into a common manifold space, and then used the features extracted from the manifold embeddings to perform domain adaptation. Nevertheless, this area of domain adaptation to handle dimensionality changes in EEG is much less explored than domain adaptation between homogeneous dimensionality \cite{gu2023generalizable}. Our work opens new ways to address the variation in input dimensionality in biosignals. Our work is one of the few studies that use GNN and domain adaptation to simultaneously address the issues of inter-subject and cross-dataset channel variations. Our work provides new solutions to enable biomedical research to integrate large disparate datasets collected over years and many different sites, which contribute to solving the common data scarcity and heterogeneity issue in medical datasets \cite{yan2020learning, yue2020deep}.

\section{Conclusion}
This work presented an end-to-end framework with GNN and transfer learning for classifying MI EEG signals using multiple datasets with different electrode layouts. We compared the binary classification accuracy of our proposed method with three different baseline setups with different trade-offs. 

Our experimental results demonstrate that the proposed network outperforms all baseline methods. This superior performance is attributed to the effective use of additional datasets without compromising non-common channels and the employment of GNNs to better handle the spatial information in the EEG layout.

In summary, our work illustrates the power of combining GNNs and transfer learning to efficiently integrate datasets without sacrificing valuable data that may not be shared across all sources. This approach offers a new way to integrate disparate biosignal databases, paving the way for novel collaboration opportunities and more comprehensive analyses in the field of biomedical research. Moreover, our findings have broader implications in the realm of machine learning in healthcare, as the developed framework can potentially be extended to other modalities and application areas where data heterogeneity and dimensionality pose significant challenges.


\section*{Acknowledgments}
We acknowledge funding from UKRI Turing AI Fellowship to AAF (EP/V025449/1).

\section*{References}
\bibliographystyle{vancouver} 
\bibliography{main.bib}

\end{document}